\documentstyle[12pt]{article}
\begin{document}

\title{Stability and Representation Dependence of the Quantum Skyrmion}
\author{A. Acus$^1$ E. Norvai\v{s}as$^1$ and D.O. Riska$^2$}
\date{}
\maketitle

\centerline{\it $^1$Institute of Theoretical Physics and Astronomy,
Vilnius, 2600 Lithuania}
\centerline{\it $^2$Department of Physics, University of Helsinki, 00014
Finland}

\begin{abstract}
A constructive realization of Skyrme's conjecture that an effective pion
mass ``may arise as a self consistent quantal effect'' based on an 
ab initio
quantum treatment of the Skyrme model is presented. In this quantum
mechanical Skyrme model the spectrum of states with $I=J$, which appears in
the collective quantization, terminates without any infinite tower of
unphysical states. The termination point depends on the 
model parameters and the dimension of the $SU(2)$ representation. 
Representations, in which the nucleon and $\Delta_{33}$ resonance are the 
only stable states, exist. The model is
developed for both irreducible and reducible representations of general
dimension. States with spin larger than 1/2 are shown to be
deformed. The representation dependence of the baryon observables
is illustrated numerically.
\end{abstract}

\vspace{1cm}

\newpage

\section{ Introduction}

\vspace{0.5cm}

The modern view of Skyrme's topological soliton model of the baryons \cite
{Sky1} is that represents a chiral symmetric effective mesonic
representation of the approximately chiral symmetric QCD Lagrangian in the
large $N_{c}$ limit, in which the baryons have to be constructed as
topological solitons \cite{Witt}. The mass of the pion, which is the
Goldstone boson of the spontaneously broken mode of chiral symmetry, is
conventionally introduced through an external chiral symmetry breaking term
in the Lagrangian density \cite{Adk}. Skyrme, however, originally suggested
a very different origin for the pion mass as ``a self consistent quantal
effect'' \cite{Sky2}. \newline

We here provide a constructive realization of this conjecture, by
demonstrating that in an ab initio quantum mechanical treatment of the
Skyrme model Lagrangian, a term, which may be interpreted as an
effective pion mass term, automatically arises. Moreover, with typical values
for the parameters in the Skyrme Lagrangian, this mass term takes values
close to that of the physical pion mass, although it is state
dependent as a consequence of the quantization procedure. This effective
pion mass may naturally be combined with the pion mass that may be
introduced by adding an explicitly chiral symmetry breaking pion mass term
to the Lagrangian density of the model.\newline

The effective pion mass appears through an additional term in the 
Euler-Lagrange equation of the quantum Skyrme model, the asymptotic
form of which is consistent with a partial conservation 
law for the axial current (PCAC). This term restores the stability of 
the soliton solutions, which is lost in a
direct variational solution of the Skyrme model, when the rotational energy
introduced by projection onto states of good spin and isospin is included 
\cite{Ban,Braa}. The quantum Skyrme model describes the
baryon states with spin larger than $1/2$ as deformed, but only
in representations with larger dimension than the fundamental
one. Finally the
spectrum of states with equal spin and isospin, which appears in the
collective quantization, terminates in the quantum Skyrme
model, and therefore unphysical states with large $I=J$ 
do not appear. The termination
point in this spectrum depends on the parameters in the model as well as on
the dimension of the representation employed. This representation
dependence is a quantum effect, which is absent in the classical 
Skyrme model \cite{Nor1}.%
\newline

The present work builds on the development of the ab initio quantized version
of the Skyrme model in ref. \cite{Nor2}, but goes beyond the treatment of
the quantum corrections as perturbations of the classical Skyrmion. The
method of ab initio treatment of the Skyrme model in $SU(2)$ representations
of arbitrary dimension is that suggested in ref. \cite{Fuji} for the
fundamental representation. The narrow stability constraints of the
fundamental representation \cite{Kost} are avoided by the development of the
model in general representations. The ab initio quantum mechanical
treatment differs from that of ref. \cite{Ceb}, in which the
systematic quantization of the Skyrme model was developed on the
classical Hamiltonian of the model. \newline

The present manuscript falls into 7 sections. In section 2 we review the
basic formalism of the Skyrme model in a representation of arbitrary
dimension. In section 3 we construct the quantum Skyrme model in
representations of arbitrary dimension. In section 4 we derive the equations
of motion and the associated effective pion mass. The Noether currents are
derived in section 5. Numerical results for the baryon observables are
presented in section 6. Section 7 contains a concluding discussion.\newline

\vspace{1cm}

\section{ The classical skyrmion in a general representation.}

\vspace{0.5cm}

The Lagrangian density of the Skyrme model \cite{Sky1} depends on a unitary
field $U(\vec r,t)$, which in a general reducible representation for the $
SU(2)$ group may be expressed as direct sum of Wigner's $D$ matrices
\begin{equation}
U(\vec r, t)=\sum_k \oplus D^{j_k}[\vec\alpha(\vec r,t)].
\end{equation}
Here $\vec\alpha$ represents a triplet of Euler angles $\alpha_1(\vec r,t),$ 
$\alpha_2(\vec r,t)$, $\alpha_3(\vec r,t)$, which form the set of the 
dynamical variables. 
The $D^{j_k}$ matrices have dimension $(2j_k+1)\times (2j_k+1)$.

The Lagrangian density has the form \cite{Nap}: 
\begin{equation}
{\cal L}[U(\vec r,t)]=-{\frac{f_\pi ^2}4} Tr\{R_\mu R^\mu \}+{\frac 1{32e^2}}
Tr\{[R_\mu ,R_\nu ]^2\},  \label{F2}
\end{equation}
where the ''right'' current $R_\mu $ is defined as 
\begin{equation}
R_\mu =(\partial _\mu U)U^{\dagger },  \label{F3}
\end{equation}
and $f_\pi $ (the pion decay constant) and $e$ are parameters.\newline

The trace of a bilinear combination of generators $\hat{J}_{a}
$ of the $SU(2)$ group depends on the representation as 
\begin{equation}
Tr\langle j{\cdot }|\hat{J}_{a}\hat{J}_{b}|j{\cdot }\rangle 
=(-)^{a}{\frac{1}{6}}\sum_{k}j_{k}(j_{k}+1)(2j_{k}+1)\delta _{a,-b}\equiv 
(-)^{a}\frac{1}{4}N\delta _{a,-b}.  \label{F5}
\end{equation}
The commutator relations for these generators are 
\begin{equation}
\lbrack \hat{J}_{a},\,\hat{J}_{b}]=\left[ 
\begin{array}{ccc}
1 & 1 & 1 \\ 
a & b & c
\end{array}
\right] \hat{J}_{c}.  \label{F6}
\end{equation}
The factor in the square brackets on the r.h.s. is the Clebsch-Gordan
coefficient $(1a1b|1c)$, in a more convenient notation. Here we have used
the normalizations $\hat{J}_{\pm }=-J_{\pm 1}/\sqrt{2}$ and $\hat{J}
_{0}=-J_{0}/\sqrt{2}$.\newline

In the classical treatment of the Skyrme model the Lagrangian density
depends on the dimension of the irreducible representation only through 
the overall scalar factor $N$ (\ref{F5}) \cite{Nor1}. In the case of a
reducible representation this overall factor is a sum over separate factors
for the different irreducible representations: 
\begin{equation}
N=2/3\sum_{k}j_{k}(j_{k}+1)(2j_{k}+1).  \label{F4}
\end{equation}
Because $N$ is an overall factor the equations of motion for 
the dynamical variables $\vec{\alpha}$ are independent of the 
dimension of the representation.\newline

The static
``spherically symmetric'' hedgehog ansatz in a general representation is
invariant under the combined spatial and isospin rotations: 
\begin{equation}
i\left[ \vec{r}\times \nabla \right] _{a}U(\vec{r})+\sqrt{2}\left[ \widehat{J
}_{a},U(\vec{r})\right] =0.  \label{F7}
\end{equation}
Here circular components are used. The solution of (\ref{F7}) 
is the generalization of the usual hedgehog ansatz 
\begin{equation}
e^{i\vec{\tau}\cdot \vec{r}F(r)}\Longrightarrow U_{0}(\vec{r})=exp\{-i\sqrt{2
}\hat{J}_{a}\cdot \hat{r}^{a}F(r)\},  \label{F8}
\end{equation}
where $\hat{r}$ is the unit vector expressed in terms of circular
components.\newline

With the hedgehog ansatz (\ref{F8}) the Lagrangian density (\ref{F2})
reduces to the following simple form 
\begin{eqnarray}
{\cal L}[F(r)] &=&-N{\cal M}(F(r))=-N\bigg\{{\frac{f_{\pi }^{2}}{2}}
\Bigl(F^{^{\prime }2}+{\frac{2}{r^{2}}}\sin ^{2}\!F\Bigr)  \nonumber \\
&&+{\frac{1}{8e^{2}}}{\frac{\sin ^{2}\!F}{r^{2}}}\Bigl(2F^{^{\prime }2}+{
\frac{\sin ^{2}\!F}{r^{2}}}\Bigr)\bigg\}.  \label{F9}
\end{eqnarray}
The requirement that the soliton mass be stationary yields the same
differential equation for the chiral angle $F(r)$ as in \cite{Nap}: 
\begin{equation}
F^{\prime \prime }+2F^{\prime \prime }\frac{\sin ^{2}F}{\widetilde{r}^{2}}
+F^{\prime 2}\frac{\sin 2F}{\widetilde{r}^{2}}+\frac{2}{\widetilde{r}^{2}}
F^{\prime }-\frac{\sin 2F}{\widetilde{r}^{2}}-\frac{\sin 2F\sin ^{2}F}{
\widetilde{r}^{4}}=0. \label{F11}
\end{equation}
Here the dimensionless variable $\widetilde{r}$ is defined as $\widetilde{r}
=ef_{\pi }r$. The boundary conditions for solitons with unit
baryon number are $F(0)=\pi$, $F(\infty )=0$. The classical soliton is 
obviously independent of the the representation.
\newline

After the renormalization the hedgehog mass in any representation has the
form:
\begin{eqnarray}
M(F) &=&\int {\rm d}^{3}r{\cal M}(F(r))=\frac{f_{\pi }}{e}\widetilde{M}(F)= 
\nonumber \\
&&2\pi \frac{f_{\pi }}{e}\int {\rm d}\tilde{r}\tilde{r}^{2}\left[ F^{\prime
2}+\frac{\sin ^{2}F}{\widetilde{r}^{2}}\left( 2+2F^{\prime 2}+\frac{\sin
^{2}F}{\widetilde{r}^{2}}\right) \right] .  \label{F10}
\end{eqnarray}

For the hedgehog solution the baryon density takes the form 
\begin{equation}
B^0=\frac{ \epsilon ^{0\nu \beta \gamma }} {24\pi^2 N} Tr\,R_\nu \,R_\beta
\,R_\gamma =-\frac 1{2\pi ^2}\frac{\sin ^2F}{r^2}F^{\prime }.  \label{F12}
\end{equation}
The renormalization factor $N$ ensures that the lowest nonvanishing baryon
number is $B=1$ for the hedgehog an in arbitrary representation.

\vspace{1cm}

\section{ Quantization in the collective coordinate approach}

\vspace{0.5cm}

The quantization of the Skyrme model in a general 
dimension \cite{Nor2} can be achieved by means of
collective rotational coordinates that separate the variables which
depend on the time and spatial coordinates \cite{Nap}: 
\begin{equation}
U(\vec r,\vec q (t))=A\left(\vec q(t)\right) U_0(\vec r) A^\dagger \left(
\vec q(t)\right) .  \label{F13}
\end{equation}
The set of three real, independent parameters $\vec
q(t)=(q^1(t),q^2(t),q^3(t))$ are quantum variables (Euler angles
representing rotations of the Skyrmion). In a general representation the
unconstrained variables $\vec q(t)$ are more convenient than the four
constrained Euler-Rodrigues parameters with the constraint used in \cite{Nap}
. When the the Skyrme Lagrangian (2) is considered quantum mechanically ab
initio the generalized coordinates $\vec q(t)$ and velocities $\vec{\dot{q}}
(t)$ satisfy the commutation relations \cite{Fuji} 
\begin{equation}
\lbrack \dot q^a,\,q^b]=-if^{ab}(\vec q).  \label{F14}
\end{equation}
Here the tensor $f^{ab}(\vec q)$ is a function of generalized coordinates $
\vec q$ only, the explicit form of which is determined after the
quantization condition has been imposed. The tensor $f^{ab}$ is symmetric
with respect to interchange of the indices $a$ and $b$ as a consequence of
the relation $[q^a,\,q^b]=0$. The commutation relation between a
generalized velocity component $\dot q^a$ and arbitrary function $G(\vec q)$
is given by 
\begin{equation}
\lbrack \dot q_a,\,G(\vec q)]=-i\sum_rf^{ar}(\vec q){\frac \partial
{\partial q^r}}G(\vec q).  \label{F15}
\end{equation}

Here Weyl ordering of the operators has been employed: 
\begin{equation}
\partial_0 G(q) ={1\over 2}\{\dot q^\alpha,{\partial\over 
\partial q^\alpha} G(q)\}.\label{X}
\end{equation}
With this choice of operator ordering no further ordering ambiguity
appears.\\
 
After making the substitution (\ref{F13}) into the Lagrangian density (\ref
{F2}) the dependence of Lagrangian on the generalized velocities can be
expressed as 
\begin{eqnarray}
L(\vec{q},\vec{\dot q},F) &=&\frac{1}{N}\int {\rm d}^{3}r{\cal 
L}(\vec{r},\vec{q}(t),F(r))=  \nonumber \\
&&-\frac{1}{4}a(F)\dot{q}^{\alpha }g(\vec{q})_{\alpha \beta }\dot{q}^{\beta
}+{\cal O}(\dot{q}^0). \label{F16}
\end{eqnarray}
Here the soliton moment of inertia $a(F)$ is given as 
\begin{eqnarray}
a(F) &=&\int {\rm d}^{3}r{\cal A}(F(r))=\frac{1}{e^{3}f_{\pi }}\tilde{a}(F)=
\nonumber \\
&&\frac{1}{e^{3}f_{\pi }}\frac{8\pi }{3}\int {\rm 
d}\tilde{r}\tilde{r}^{2}\sin ^{2}F\left[ 1+F^{\prime 2}+\frac{\sin 
^{2}F}{\widetilde{r}^{2}}\right] ,  \label{F17}\end{eqnarray}
where ${\cal A}(F(r))$ is the moment of inertia density.

The 3$\times $3 metric tensor $g(\vec{q})_{\alpha \beta }$ is defined as the
scalar product of a set of functions $C_{\alpha }^{(m)}(\vec{q})$ \cite{Nor1}
\begin{eqnarray}
g(\vec{q})_{\alpha \beta } &=&\sum_{m}(-)^{m}C_{\alpha }^{(m)}C_{\beta
}^{(-m)}=\sum_{m}(-)^{m}C_{\alpha }^{\prime (m)}C_{\beta }^{\prime (-m)}= 
\nonumber \\
&&-2\delta _{\alpha \beta }-2(\delta _{\alpha 1}\delta _{\beta 3}+\delta
_{\alpha 3}\delta _{\beta 1})\cos q^{2}.  \label{F18}
\end{eqnarray}
Here the functions $C_{\alpha }^{^{\prime }(m)}$ are defined in \cite{Nor2}
and related with $C_{\alpha }^{(m)}$ as 
\begin{equation}
C_{\alpha }^{(m)}(\vec{q})=\sum_{m}D_{m,m^{\prime }}^{1}(\vec{q})C_{\alpha
}^{^{\prime }(m)}(\vec{q}).
\end{equation}
The canonical momentum $p_{\alpha }$, which is conjugate to $q^{\alpha }$,
is defined as 
\begin{equation}
p_{\alpha }(\vec{q},\vec{\dot q},F)=\frac{\partial 
L(\vec{q},\vec{\dot q},F)}{\partial \dot{q}^{\alpha 
}}=-\frac{1}{4}a(F)\{\dot{q}^{\beta },g(\vec{q})_{\beta\alpha }\},  
\label{F19}
\end{equation}
where the curly bracket denotes an anticommutator. The canonical commutation
relations 
\begin{equation}
\left[ p_{\alpha }(\vec{q},\vec{\dot q},F),q^{\beta }\right] =-i\delta 
_{\alpha\beta },  \label{F20}
\end{equation}
then yield the following explicit form for the functions $f^{ab}(\vec{q})$: 
\begin{equation}
f^{ab}(\vec{q})=-\frac{2}{a(F)}g_{\alpha \beta }^{-1}(\vec{q}).  \label{F21}
\end{equation}

Because of the nonlinearity of the Skyrme model the canonical momenta
defined in this way do not necessarily satisfy the relation $[p_{\alpha
},p_{\beta }]=0$. As shown in \cite{Fuji}, there exists a local
transformation of the set of variables $\vec{q}$, which makes it possible to
satisfy these relations. Following Fujii et al. \cite{Fuji} we define an
angular momentum operator: 
\begin{equation}
\hat{J}_{a}^{\prime }=-\frac{i}{2}\left\{ p_{r},C_{\left( -a\right)
}^{\prime r}(\vec{q})\right\} =(-)^{a}\frac{ia(F)}{4}\left\{ \dot{q}
^{r},C_{r}^{\prime \left( -a\right) }(\vec{q})\right\} ,  \label{F22}
\end{equation}
which satisfies the commutation relation (\ref{F6}). It is readily
verified 
that the operator $\hat{J}_{a}^{\prime }$ is a $D^{j}(\vec{q})$ ''right
rotation'' generator. The explicit form for the Lagrangian of the
consistently quantized Skyrme model now takes the form: 
\begin{eqnarray}
L(\vec{q},\vec{\dot q},F) &=&-M(F)-\Delta M_{\Sigma j}(F)+\frac{1}{a(F)}\hat{
J^{\prime }}^{2}=  \nonumber \\
&&\ -M(F)-\Delta M_{\Sigma j}(F)+\frac{1}{a(F)}\hat{J}^{2},  \label{F23}
\end{eqnarray}
where 
\begin{eqnarray}
&&\Delta M_{\Sigma j}(F)=\int {\rm d}^{3}r\Delta {\cal M}_{\Sigma
j}(F(r))=e^{3}f_{\pi }\cdot \Delta \widetilde{M}_{\Sigma j}(F)=-e^{3}f_{\pi }
\frac{2\pi }{15\tilde{a}^{2}(F)}  \nonumber \\
&&\times \int {\rm d}\tilde{r}\tilde{r}^{2}\sin ^{2}F\Bigl\{15+4d_{2}\sin
^{2}F(1-F^{\prime 2})+2d_{3}\frac{\sin ^{2}F}{\tilde{r}^{2}}+2d_{1}F^{\prime
2}\Bigr\}.  \label{F24}
\end{eqnarray}

The coefficients $d_{j}$ in these expressions are given as 
\begin{eqnarray}
d_{1} &=&\frac{1}{N}\sum_{k}j_{k}(j_{k}+1)(2j_{k}+1)\left[
8j_{k}(j_{k}+1)-1\right] ,  \label{F24b} \\
d_{2} &=&\frac{1}{N}\sum_{k}j_{k}(j_{k}+1)(2j_{k}+1)(2j_{k}-1)(2j_{k}+3),
\label{F24c} \\
d_{3} &=&\frac{1}{N}\sum_{k}j_{k}(j_{k}+1)(2j_{k}+1)\left[
2j_{k}(j_{k}+1)+1\right] .  \label{F24d}
\end{eqnarray}
The corresponding Hamilton operator is 
\begin{equation}
H_{j}(F)=M(F)+\Delta M_{j}(F)+\frac{1}{a(F)}\hat{J^{\prime }}
^{2}=M(F)+\Delta M_{j}(F)+\frac{1}{a(F)}\hat{J}^{2}.  \label{F25}
\end{equation}

This result differs from the semiclassical one in the appearance of the
negative quantum correction $\Delta M_j(F)$ \cite{Fuji}, which depends on
the dimension of the representation of the $SU(2)$ group \cite{Nor2}.\newline

For the Hamiltonian (\ref{F25}) the normalized state vectors with fixed spin
and isospin $\ell $ are 
\begin{equation}
\left| 
\begin{array}{c}
\ell  \\ 
{m,m^{\prime }}
\end{array}
\right\rangle =\frac{\sqrt{2\ell +1}}{4\pi }D_{m,m^{\prime }}^{\ell }(\vec{q}
)\left| 0\right\rangle ,  \label{F26}
\end{equation}
with the eigenvalues 
\begin{equation}
H(j,\ell ,F)=M(F)+\Delta M_{j}(F)+\frac{\ell (\ell +1)}{2a(F)}.  \label{F27}
\end{equation}

Substitution of the rotated hedgehog (\ref{F13}) into the Lagrangian density
(\ref{F2}) yields the following expression for the 
Lagrangian density for the quantum Skyrme model in a
general reducible representation: 
\begin{equation}
{\cal L}(\vec{r},\vec{q}(t),F(r))=\frac{3{\cal A}(F(r))}{2\,a^{2}(F)}\left( 
\widehat{J}^{\prime 2}-(\widehat{J}^{\prime }\cdot \hat{r})(\widehat{J}
^{\prime }\cdot \hat{r})\right) -\Delta {\cal M}_{\Sigma j}(F(r))-{\cal M}
(F(r)). \label{F41}
\end{equation}
The angular momentum operator on the r.h.s. of (\ref{F41}) can be separated
into scalar and tensor terms in the usual way: 
\begin{eqnarray}
\widehat{J}^{\prime 2}-(\widehat{J}^{\prime }\cdot \hat{r})(\widehat{J}
^{\prime }\cdot \hat{r}) &=&\frac{2}{3}\widehat{J}^{\prime
2}-\frac{4\pi }{3}Y_{2,m+m^{\prime }}^*(\vartheta ,\varphi
)  \nonumber \\
&&\times \left[ 
\begin{array}{ccc}
1 & 1 & 2 \\ 
m & m^{\prime } & m+m^{\prime }
\end{array}
\right] \widehat{J}_{m}^{\prime }\widehat{J}_{m^{\prime }}^{\prime },
\label{F42}
\end{eqnarray}
where $Y_{l,m}(\vartheta ,\varphi )$ is a spherical harmonic.

The volume integral of the Lagrangian density (\ref{F41}) 
gives the Lagrangian (\ref{F23}).
In the fundamental representation, for which $j=1/2$, the 
second rank tensor part of (\ref{F42}) vanishes. This implies that 
the quadrupole moment of the
$\Delta_{33}$ resonance cannot be described with the fundamental 
representation. \newline

The Hamiltonian density, which corresponds to the quantum Lagrangian (\ref
{F41}) has the following matrix elements for baryon states with spin
and isospin $\ell >1/2$: 
\[
\left\langle 
\begin{array}{c}
\ell \\ 
m_{t}m_{s}
\end{array}
\right| {\cal H}(\vec{r},\vec{q}(t),F(r))\left| 
\begin{array}{c}
\ell \\ 
m_{t}m_{s}
\end{array}
\right\rangle ={\cal M}(F(r))+\Delta {\cal M}_{\Sigma j}(F(r)) 
\]
\begin{equation}
+\frac{{\cal A}(F(r))}{2\,a^{2}(F)}\left( \ell (\ell +1)-\sqrt{\frac{2}{3}}
\pi [3m_{s}^{2}-\ell (\ell +1)]Y_{2,0}(\vartheta ,\varphi )\right) .
\label{F43}
\end{equation}
For nucleons $\ell =1/2$ the dependence on angles are absent and the quantum
skyrmion is therefore spherically symmetric as required. 

\vspace{1cm}

\section{ The Chiral Angle and the Pion Mass}

\vspace{0.5cm}

The $I=J=\ell =1/2$ and $I=J=\ell =3/2$ skyrmions are to be identified with
the nucleons and the $\Delta _{33}$ resonances. Minimization of the
classical expression for the mass $M(F)$ (\ref{F10}) leads to the
conventional differential equation for the chiral angle $F(r)$ (\ref{F11}
) according to which $F(r)$ falls as $1/{r}^{2}$ at large distances.
\newline

In the semiclassical approach the quantum mass term $\Delta {M}
_{\Sigma j}$ is absent from the mass expression (\ref{F25}). Its
absence has the
consequence that variation of the truncated quantum mass expression
yields no stable solution \cite{Ban,Braa}. The
semiclassical approach describes the skyrmion as a ``rotating'' rigid-body
with fixed $F(r)$ \cite{Nap}. In contrast the full energy expression (\ref
{F25}) that is obtained in the consistent canonical quantization procedure
in collective coordinates approach gives stable solutions.\newline

Minimization of the quantum mass expression (\ref{F25}), leads to the
following integro-differential equation for the chiral angle $F(r)$: 
\begin{eqnarray}
&&F^{\prime \prime }\Bigl[-2\tilde{r}^{2}-4\sin ^{2}F+\frac{e^{4}\tilde{r}
^{2}\sin ^{2}F}{15\tilde{a}^{2}(F)}\left\{ 80\tilde{a}(F)\Delta \widetilde{M}
_{\Sigma j}(F)+20\ell (\ell +1)\right.  \nonumber \\
&&\left. +4d_{1}-8d_{2}\sin ^{2}F\right\} \Bigr]+F^{\prime 2}\Bigl[-2\sin 2F+
\frac{e^{4}\tilde{r}^{2}\sin 2F}{15\tilde{a}^{2}(F)}\left\{ 40\tilde{a}
(F)\Delta \widetilde{M}_{\Sigma j}(F)\right.  \nonumber \\
&&\left. \left. +10\ell (\ell +1)+2d_{1}-8d_{2}\sin ^{2}F\right\} \right]
+F^{\prime }\Bigl[-4\tilde{r}+\frac{e^{4}\tilde{r}\sin ^{2}F}{15\tilde{a}
^{2}(F)}  \nonumber \\
&&\left. \times \left\{ 160\tilde{a}(F)\Delta \widetilde{M}_{\Sigma
j}(F)+40\ell (\ell +1)+8d_{1}-16d_{2}\sin ^{2}F\right\} \right] +\sin
2F\Bigl[2  \nonumber \\
&&+2\frac{\sin ^{2}F}{\tilde{r}^{2}}-\frac{e^{4}}{15\tilde{a}^{2}(F)}\left\{
\left( 40\tilde{a}(F)\Delta \widetilde{M}_{\Sigma j}(F)+10\ell (\ell
+1)\right) \left( \tilde{r}^{2}+2\sin ^{2}F\right) \right.  \nonumber \\
&&\left. \left. +15\tilde{r}^{2}+4d_{3}\sin ^{2}F+8d_{2}\tilde{r}^{2}\sin
^{2}F\right\} \right] =0,  \label{f9}
\end{eqnarray}
with the usual boundary conditions $F(0)=\pi $ and $F(\infty )=0$.\newline
The state dependence of this equation of motion is a direct
consequence of the fact that quantization preceeded variation
(cf. ref. \cite{Liu}).

At large distances this equation reduces to the asymptotic form: 
\begin{equation}
\tilde{r}^{2}F^{\prime \prime }+2\tilde{r}F^{\prime }-(2+\tilde{m}_{\pi }^{2}
\tilde{r}^{2})F=0,  \label{f10}
\end{equation}
where the quantity $\tilde{m}_{\pi }^{2}$ is defined as 
\begin{equation}
\tilde{m}_{\pi }^{2}=-\frac{e^{4}}{3\tilde{a}(F)}\left\{ 8\Delta \widetilde{M
}_{\Sigma j}(F)+\frac{2\ell (\ell +1)+3}{\tilde{a}(F)}\right\} .  \label{f11}
\end{equation}
The corresponding asymptotic solution takes the form 
\begin{equation}
F(\tilde{r})=k\left( \frac{\tilde{m}_{\pi }}{\tilde{r}}+\frac{1}{\tilde{r}
^{2}}\right) \exp (-\tilde{m}_{\pi }\tilde{r}).  \label{f12}
\end{equation}
The requirement of stability of the quantum skyrmion is that the integrals   
(\ref{F10}), (\ref{F17}) and (\ref{F24}) converge. This requirement is
satisfied only if $\tilde{m}_{\pi }^{2}>0$. For that the presence of the
negative quantum correction $\Delta {M_{j}}(F)$ is necessary. It is
the absence of this term, which leads to the instability of the skyrmion 
in the
semiclassical approach \cite{Braa}. Note that in the quantum treatment 
the chiral angle
has the asymptotic Yukawa behaviour (\ref{f12}) even in the chiral limit 
\cite{Fuji}.
The positive quantity $m_{\pi }=ef_{\pi }\tilde{m}_{\pi }$ should obviously
be interpreted as an effective mass for the pion field in
the Skyrmion mass.\\

In contrast to the classical skyrmion the stability of the quantum
mechanical skyrmion depends on the Lagrangian parameter values $f_{\pi }$
and $e$ \cite{Kost}. Moreover positivity of the pion mass (\ref{f11}),
can obviously only be achieved for states with sufficiently small values of
spin $\ell $. This implies that the spectrum of states with equal spin and
isospin will necessarily terminate at some finite value of the spin
quantum number. As the negative quantum mass correction $\Delta {M}
_{\Sigma j}$ in the expression (\ref{F24}) grows in magnitude with the
dimension of the representation, it is always possible to find a
representation in which the nucleon and the $\Delta _{33}$ resonance are the
only stable particles, as required by experiment. This argument 
is more general than 
the method of self consistent dynamical truncation of the spectrum 
suggested in ref. \cite{Blai}. 

\vspace{1cm}

\section{ The Noether currents}

\vspace{0.5cm}

The Lagrangian density of the Skyrme model is invariant under left and right
transformations of the unitary field $U$. The corresponding Noether currents
can be expressed in terms of the collective coordinates (\ref{F13}). The
vector and axial Noether currents that are associated with the
transformations,

\begin{equation}
U(x)\stackrel{V(A)}{\longrightarrow }\left( 1-i2\sqrt{2}\omega ^a\hat
J_a\right) U(x)\left( 1+(-)i2\sqrt{2}\omega ^a\hat J_a\right),  \label{F28}
\end{equation}
are nevertheless simpler and directly related to physical observables. The
factor $-2\sqrt{2}$ before generators is needed in the case $j=1/2$ to match
the transformation (\ref{F28}) with the infinitesimal transformation in \cite
{Nap}. The Noether currents are operators that depend on the generalized
collective coordinates $\vec q$ and the generalized angular momentum 
operators $\hat J^{\prime }_a$ (\ref{F22}). \\

The explicit expression for the spatial components of the vector
current density is 
\begin{eqnarray}
\hat V_b^a&=&\frac{\partial {\cal L}_V}{\partial \left( \nabla ^b\omega
_a\right) }=2\sqrt{2}\frac{\sin ^2F}r\Biggl(i\biggl\{ f_\pi ^2+\frac
1{e^2}\Bigl(F^{\prime }{}^2  \nonumber \\
&&+\frac{\sin ^2F}{r^2}-\frac{2d_2+5}{4\cdot 5\cdot a^2(F)}\sin ^2F\Bigr) 
\biggr\} \left[ 
\begin{array}{ccc}
1 & 1 & 1 \\ 
u & s & b
\end{array}
\right] D_{a,s}^1(\vec q)\hat x_u  \nonumber \\
&&-\frac{\sin ^2F}{\sqrt{2}\cdot e^2\cdot a^2(F)}(-)^s\Bigl\{\left[ \hat
J^{\prime }\times \hat r\right] _{-s}D_{a,s}^1(\vec q)\left[ \left[ \hat
J^{\prime }\times \hat r\right] \times \hat r \right] _b  \nonumber \\
&&+\left[ \left[ \hat J^{\prime }\times \hat r\right] \times \hat r \right]
_bD_{a,s}^1(\vec q)\left[ \hat J^{\prime }\times \hat r\right] _{-s}\Bigr\}
\Biggr).  \label{F29}
\end{eqnarray}
Here $\nabla ^k$ is a circular component of the gradient operator. The
indices $a$ and $b$ denote isospin and spin components. The time (charge)
component of the vector current density has the expression
\begin{eqnarray}
\hat V_t^a&=&\frac{\partial {\cal L}_V}{\partial \left( \partial _0\omega
_a\right) }=\frac{2\sqrt{2}}{a(F)}\sin ^2F\left[ f_\pi ^2 +\frac
1{e^2}\left( F^{\prime 2}+\frac{\sin ^2F}{r^2}\right) \right]  \nonumber \\
&&\times (-)^s\left\{ D_{a,-s}^1(\vec q)\hat J_s^{\prime }-D_{a,-s}^1(\vec q
)\hat x_s(\hat J^{\prime }\cdot \hat r)\right\}.  \label{F30}
\end{eqnarray}
The explicit expression for the axial current density takes the form 
\begin{eqnarray}
&&\hat A_b^a=\frac{\partial {\cal L}_A}{\partial \left( \nabla ^b\omega
_a\right) }=\Biggl( \Biggl\{ f_\pi ^2\frac{\sin 2F}r+\frac 1{e^2}\frac{\sin
2F}r\biggl( F^{\prime }{}^2+\frac{\sin ^2F}{r^2}-\frac{\sin ^2F}{4\cdot
a^2(F)}\biggr) \Biggr\}  \nonumber \\
&&\times D_{a,b}^1(\vec q)+\Biggl\{ f_\pi ^2\Bigl(2F^{\prime }-\frac{ \sin 2F
}r\Bigr)-\frac 1{e^2}\biggl(F^{\prime }{}^2\frac{\sin 2F}r-4F^{\prime }\frac{
\sin ^2F}{r^2}  \nonumber \\
&&+\frac{\sin ^2F\sin 2F}{r^3}-\frac{\sin ^2F\sin 2F}{4\cdot a^2(F)\cdot r}
\Bigr)\Biggr\} (-)^sD_{a,s}^1(\vec q)\hat x_{-s}\hat x_b-\frac{ 2F^{\prime
}\sin ^2F}{e^2\cdot a^2(F)}  \nonumber \\
&&\times (-)^s\left\{ D_{a,s}^1(\vec q)\hat x_{-s}\hat J^{\prime }{}^2+\hat
J^{\prime }{}^2D_{a,s}^1(\vec q)\hat x_{-s}-2D_{a,s}^1(\vec q)\hat
x_{-s}(\hat J^{\prime }\cdot \hat r)(\hat J^{\prime }\cdot \hat r)\right\}
\hat r_b  \nonumber \\
&&-\frac{\sin ^2F\sin 2F}{e^2\cdot a^2(F)\cdot r}(-)^s\left\{ \left[ \left[
\hat J^{\prime }\times \hat r\right] \times \hat r \right]
_{-s}D_{a,s}^1(\vec q)\left[ \left[ \hat J^{\prime }\times \hat r \right]
\times \hat r\right] _b\right.  \nonumber \\
&&\left. +\left[ \left[ \hat J^{\prime }\times \hat r\right] \times \hat
r\right] _bD_{a,s}^1(\vec q)\left[ \left[ \hat J^{\prime }\times \hat
r\right] \times \hat r\right] _{-s}\right\} \Biggr).  \label{F31}
\end{eqnarray}
The operators (\ref{F29})-(\ref{F31}) are well defined for all
representations $j$ of the classical soliton and for fixed spin and isospin $
\ell $ of the quantum skyrmion. The new terms, which are absent in the
corresponding semiclassical expression, are those that have the 
factor $a^2(F)$ in the denominator.\\

The conserved topological current density in Skyrme model is the baryon
current density, the components of which are 
\begin{equation}
{\cal B}_a(\vec r ,F(r))=\frac 1{\sqrt{2}\pi ^2a(F)\ r}\sin ^2F\cdot
F^{\prime }\left[ \hat J^{\prime }\times \hat r \right] _a.  \label{F32}
\end{equation}

The matrix elements of the third component of the isoscalar
magnetic moment operator have the form 
\begin{eqnarray}
\left\langle 
\begin{array}{c}
\ell \\ 
m_tm_s
\end{array}
\right| \left[ \mu _{I=0}\right] _3\left| 
\begin{array}{c}
\ell \\ 
m_tm_s
\end{array}
\right\rangle &=&\left\langle 
\begin{array}{c}
\ell \\ 
m_tm_s
\end{array}
\right| \frac 12\int d^3xr\left[ \hat r\times {\cal B} \right] _0\left| 
\begin{array}{c}
\ell \\ 
m_tm_s
\end{array}
\right\rangle =  \nonumber \\
&&\frac{\left[ \ell (\ell +1)\right] ^{1/2}e}{3\cdot \tilde a(F)f_\pi }
\left\langle \tilde r_{I=0}^2\right\rangle \left[ 
\begin{array}{ccc}
\ell & 1 & \ell \\ 
m_s & 0 & m_s
\end{array}
\right] .  \label{F33}
\end{eqnarray}
Here the isoscalar mean square radius is given as 
\begin{equation}
\left\langle r_{E,I=0}^2\right\rangle =\frac 1{e^2f_\pi ^2}\left\langle
\tilde r_{I=0}^2\right\rangle =-\frac 1{e^2f_\pi ^2}\frac 2\pi \int \tilde
r^2\sin ^2F\cdot F^{\prime }d\tilde r, \label{F34}
\end{equation}
and the quantity  $\tilde a$ is defined in eq. (\ref{F17}).\\

The matrix elements of the third component of the isovector part of magnetic
moment operator that is obtained from the vector current density (\ref{F29})
have the form 
\begin{eqnarray}
&&\left\langle 
\begin{array}{c}
\ell \\ 
m_{t}m_{s}
\end{array}
\right| \left[ \mu _{I=1}\right] _{3}\left| 
\begin{array}{c}
\ell \\ 
m_{t}m_{s}
\end{array}
\right\rangle =\left\langle 
\begin{array}{c}
\ell \\ 
m_{t}m_{s}
\end{array}
\right| \frac{1}{2}\int d^{3}x\cdot r\left[ \hat{r}\times \hat{V}^{3}\right]
_{0}\left| 
\begin{array}{c}
\ell \\ 
m_{t}m_{s}
\end{array}
\right\rangle =  \nonumber \\
&&\ \left[ \frac{\tilde{a}(F)}{e^{3}\cdot f_{\pi }}+\frac{8\pi \cdot e}{
3\cdot f_{\pi }\cdot \tilde{a}^{2}(F)}\int d\tilde{r}\cdot \tilde{r}^{2}\sin
^{4}F\right. \left( 1-\frac{d_{2}}{2\cdot 5}\right. -\frac{\ell (\ell +1)}{3}
\nonumber \\
&&\left. \left. +\frac{(-)^{2\ell }}{2}{\Biggl[\frac{5\ell (\ell +1)(2\ell
-1)(2\ell +1)(2\ell +3)}{2\cdot 3}\Biggr]}^{1/2}\left\{ 
\begin{array}{ccc}
1 & 2 & 1 \\ 
\ell & \ell & \ell
\end{array}
\right\} \right) \right]  \nonumber \\
&&\times \left[ 
\begin{array}{ccc}
\ell & 1 & \ell \\ 
m_{s} & 0 & m_{s}
\end{array}
\right] \left[ 
\begin{array}{ccc}
\ell & 1 & \ell \\ 
m_{t} & 0 & m_{t}
\end{array}
\right] .  \label{F35}
\end{eqnarray}
Here the symbol in the curly brackets is a $6j$ coefficient.

The volume integral of the axial current density (42) yields the 
axial coupling constant $g_A$
as 
\begin{equation}
g_A=-3\left\langle 
\begin{array}{c}
1/2 \\ 
{1/2,1/2}
\end{array}
\right| \int d^3xA_0^0\left| 
\begin{array}{c}
1/2 \\ 
{1/2,1/2}
\end{array}
\right\rangle =\frac 1{e^2}\tilde {g_1}(F)-\frac{\pi ^2e^2}{3\cdot \tilde
a^2(F)}\left\langle \tilde r_{I=0}^2\right\rangle ,  \label{F36}
\end{equation}
where 
\begin{eqnarray}
\tilde g_1(F)&=&\frac{4\pi }3\int d\tilde r(\tilde r^2F^{\prime }+\tilde
r\sin 2F+\tilde r\sin 2F\cdot F^{\prime 2}  \nonumber \\
&&+2\sin ^2F\cdot F^{\prime }+\frac{\sin ^2F}{\tilde r}\sin 2F).  \label{F37}
\end{eqnarray}

For nucleons the the isovector charge mean square radius becomes
\begin{equation}
\left\langle r_{E,I=1}^2\right\rangle =\frac 1{e^2f_\pi ^2}\left\langle
\tilde r_{E,I=0}^2\right\rangle =\frac 1{e^2f_\pi ^2}\frac{\int d \tilde
r\tilde r^4\sin ^2F\left[ 1+F^{\prime 2}+\frac{\sin ^2F}{\tilde r} \right] }{
\int d\tilde r\tilde r^2\sin ^2F\left[ 1+F^{\prime 2}+ \frac{\sin ^2F}{
\tilde r}\right] }.  \label{F38}
\end{equation}

The isoscalar magnetic mean square radius has the
expression
\begin{equation}
\left\langle r_{M,I=0}^2\right\rangle =\frac 1{e^2f_\pi ^2}\left\langle
\tilde r_{M,I=0}^2\right\rangle =-\frac 1{e^2f_\pi ^2}\frac 2\pi \frac{\int
d\tilde r\tilde r^4\sin ^2F\cdot F^{\prime }}{\int d\tilde r\tilde r^2\sin
^2F\cdot F^{\prime }},  \label{F39}
\end{equation}
and the isovector magnetic mean square radius the expression 
\begin{eqnarray}
\left\langle r_{M,I=1}^2\right\rangle &=&\frac 1{e^2f_\pi ^2}\left\langle
\tilde r_{M,I=0}^2\right\rangle =  \nonumber \\
&&\frac 1{e^2f_\pi ^2}\frac{\int d\tilde r\tilde r^4\sin ^2F\left[
1+F^{\prime 2}+\frac{\sin ^2F}{\tilde r}+\frac{e^2\sin ^2F}{\tilde a^2(F)}
\left( \frac 34-\frac{d_2}{10}\right) \right] }{\int d\tilde r\tilde r^2\sin
^2F\left[ 1+F^{\prime 2}+\frac{\sin ^2F}{\tilde r}+\frac{ e^2\sin ^2F}{
\tilde a^2(F)}\left( \frac 34-\frac{d_2}{10}\right) \right] }.  \label{F40}
\end{eqnarray}

The matrix element of the divergence of the vector current (\ref{F29}) 
vanishes:
\begin{equation}
\left\langle 
\begin{array}{c}
\ell \\ 
m_{t}m_{s}
\end{array}
\right| \nabla ^{b}\widehat{V}_{b}^{a}\left| 
\begin{array}{c}
\ell \\ 
m_{t}m_{s}
\end{array}
\right\rangle =0.  \label{f13}
\end{equation}

The the asymptotic equation of motion (\ref{f10}), valid for large r,
is recovered if matrix element of the divergence of the axial 
current (\ref{F31}) for the proton is taken to be
\begin{equation}
\left\langle 
\begin{array}{c}
\ell \\ 
m_{t}m_{s}
\end{array}
\right| \nabla ^{b}\widehat{A}_{b}^{a}\left| 
\begin{array}{c}
\ell \\ 
m_{t}m_{s}
\end{array}
\right\rangle =f_{\pi }^{2} m_{\pi }^{2}F(r).  \label{f14}
\end{equation}
This is the usual equation for a ``partially conserved
axial current'' (PCAC), and supports the interpretation of
$m_\pi$ as an effective pion mass.\\

\section{Numerical Results}
The equation of motion for the quantum Skyrme model (\ref{f9}) 
depends -- in contrast to the classical case -- on the parameter $
e$ and representation. Moreover the differential equation for the
chiral angle is state dependent.\\

For nucleons ($\ell=1/2$) solutions for the chiral
angle, which describe stable solitons with spin $1/2$,
exist when $e<7.5$. The largest value of $e$, for which stable
solutions are obtained, decreases with increasing dimensionality of the
representation. For $\Delta _{33}$ resonances ($\ell=3/2$) there are
no stable soliton solutions in the fundamental representation, nor
in the representation with $j=1$ in the quantum 
Skyrme model. In the representations with
$j=3/2$ and $2$ there are only stable soliton solutions for baryons
with spin $\ell=1/2$ and $3/2$. A dimension with $j=5/2$ allows
stable solitons with spin $\ell=1/2$, $3/2$ and $5/2$, and therefore
appears to be empirically contraindicated.\\

The two parameters of the 
model, $f_{\pi }$ and $e$, may be determined in the
usual way by fitting two empirical baryon observables.. The procedure
adopted here was to first determine these two parameters 
by using the chiral angle of the classical Skyrme model, which
is independent of both the model parameters and the dimension of
the representation \cite{Nor1}, by a fit to the nucleon mass (939 MeV) 
and its
isoscalar radius (0.72 fm) for different values of the
dimension $j$ of the representation. These parameters where then used
in a numerical solution of the equation (\ref{f9}). That solution
was subsequently used to determine new values of $f_{\pi }$ and $e$. This
procedure was iterated until a converged solution was obtained. The
numerical results are shown in Table 1. For the irreducible 
representation with $j=1$ the proton magnetic moment calculated
in this way is within 10\%  of the empirical value. The
calculated values of both the neutron magnetic moment and the axial 
coupling constant agree with the corresponding empirical values
to within 1\%. The $\Delta _{33}$ resonance observables for different 
representations as obtained with fixed values for $f_{\pi }$ and $e$
are presented in Table 2.\\

\section{Discussion}

There are two main aspects of the quantum corrections to the Skyrme
model based description of the baryons. One is the treatment of the
dynamical field variables of the Lagrangian density as quantum mechanical
variables ab initio. This very likely formed the basis for
Skyrme's conjecture for the origin of the pion mass \cite{Sky2}.
The development of the ab initio quantum mechanical treatment of
the model was pioneered in ref. \cite{Fuji}, and was developed
above to realize Skyrme's conjecture constructively. The other 
main aspect is
the treatment of quantum fluctuations of the pion field as loop
corrections \cite{Mei}.\\

Both types of quantum effects lead to substantial modifications
of the phenomenological description of the baryons based on
the Skyrme model. Both also lead to negative quantum corrections
to the Skyrmion mass. In the present work this negative
mass correction was shown to imply a positive effective mass for
the pion field in the Skyrmion, and to stable variational
solutions for the quantum soliton.\\

The spectrum of states with $I=J$ terminates in the quantum
Skyrme model, because the effective pion mass becomes 
negative for sufficiently large spin. Moreover it describes
the states with spin larger than $1/2$ as deformed. That
deformation is inherent in a Skyrmion model with a
terminating spectrum has also been been noted in \cite{Dor},
although in that work the absence of states with large
spin and isospin was achieved by associating very large
decay widths with those states.\\

The systematic quantum mechanical treatment of the Skyrmion was
shown to imply the need to employ representations of larger
dimension than the fundamental one for the description of the nucleon
and the $\Delta_{33}$ resonance as stable solitons with spin and
isospin $1/2$ and $3/2$ respectively in the same representation. 
The quantum treatment 
implies that the tower of states with $I=J$ terminates, and that
there therefore is no infinite tower of unphysical states as
in the semiclassical approach. 



\bibliographystyle{plain}

\newpage

\begin{table}
\caption{The predicted static nucleon observables in different
representations with fixed empirical values for the isoscalar
radius $0.72$ fm. and nucleon mass $939$ MeV.}

\label{table2}
\begin{tabular}{c|c|c|c|c|c|c}
\hline
$j$&$1/2$&$1$&$3/2$ 
&$5/2$&$\scriptscriptstyle 1\oplus \frac{1}{2} \oplus \frac 
{1}{2}$&\rm{Exp.} \\ \hline$m_N$ 
&input&input&input&input&input&$939$ MeV \\ \hline
$f_\pi $& 59.8 & 58.5 & 57.7 & 56.6 & 58.8 &$93$ 
MeV \\\hline
$e$ & 4.46 & 4.15 & 3.86 &  3.41 & 4.24 &  \\ 
\hline
$\mu_p$ & 2.60 & 2.52 & 2.51 & 2.52 & 2.53 &$2.79$\\
\hline 
$\mu_n$ & $-$2.01 & $-$1.93 & $-$1.97 & $-$2.05 & 
$-$1.93 & $-1.91$ \\
\hline
$g_A$ &1.20&1.25&1.33&1.52&1.23&$1.26$\\
\hline
$m_\pi $& 79.5 & 180. & 248. & 336. & 155.&$138$MeV 
\\ \hline
$\sqrt{\langle r^2_E\rangle }_{I=0}$ 
&input&input&input&input&input&$0.72$fm \\
\hline
$\sqrt{\langle r^2_E\rangle }_{I=1}$ 
&1.33&1.03&0.97&0.93&1.07&$0.88$fm\\ 
\hline
$\sqrt{\langle r^2_M\rangle }_{I=0}$ 
&1.05&1.01&1.00&1.00&1.01&$0.81$fm \\ 
\hline
$\sqrt{\langle r^2_M\rangle }_{I=1}$ 
&1.32&1.03&0.97&0.93&1.07&$0.80$fm \\ 
\hline
\end{tabular}
\end{table}

\begin{table}
\caption{The predicted static $\Delta _{33}$-resonances observables in 
different representations with fixed values for the 
parameters $e=4.15$ and $f_\pi =58.5$ (determined by a fit to the
nucleon observables $m_N =939$, 
$\langle r^2\rangle ^{1/2}_{I=0}=0.72$, in a representation with $j=1$).}

\label{tablerez7}
\begin{tabular}{c|c|c|c|c}
\hline
$j$ &$\scriptscriptstyle \frac {3}{2} \oplus 1\oplus \frac 
{1}{2}$ & $\frac {3}{2}$ & $2$  &\rm Exp. \\ 
\hline$m_\Delta $ & 
1055. & 1029. & 910. &$1232$ MeV \\ \hline
$\mu_{\Delta ^{++}}$ & 7.38 & 6.40 & 4.20 &$3.7-7.5$\\
\hline 
$\mu_{\Delta ^{+}}$& 3.02 & 2.73 & 2.01 &?\\
\hline 
$\mu_{\Delta ^{0}}$ & $-$1.33 &$-$0.94 & $-$0.19 &?\\
\hline 
$\mu_{\Delta ^{-}}$& $-$5.69 & $-$4.61 & $-$2.38 & ? \\
\hline
$\sqrt{\langle r^2_E\rangle }_{I=0}$ 
&0.91&0.87&0.72&? \\
\hline
\end{tabular}
\end{table}

\end{document}